\documentclass[conference]{IEEEtran}
\IEEEoverridecommandlockouts
\usepackage{cite}
\usepackage{hyperref}
\usepackage{amsmath,amssymb,amsfonts}
\usepackage{algorithmic}
\usepackage{graphicx}
\usepackage{nameref}
\usepackage{textcomp}
\usepackage{xcolor}
\usepackage{multirow}
\usepackage{graphicx}

\usepackage[utf8]{inputenc}
\usepackage{booktabs}
\usepackage{subfig}
\usepackage{xspace}
\usepackage{siunitx}
\usepackage{listings}
\usepackage{amsmath,amssymb,amsfonts}
\usepackage{algorithmic}

\def\BibTeX{{\rm B\kern-.05em{\sc i\kern-.025em b}\kern-.08em
    T\kern-.1667em\lower.7ex\hbox{E}\kern-.125emX}}
\usepackage{hyperref}

\def\eg{\emph{e.g. }\xspace}
\def\etc{\emph{etc. }\xspace}

\newcommand{\pb}[1]{\vspace{0.75ex}\noindent{\bf \em #1}\hspace*{.3em}}

\newcommand{\one}{({\em i}\/)}
\newcommand{\two}{({\em ii}\/)}
\newcommand{\three}{({\em iii}\/)}
\newcommand{\four}{({\em iv}\/)}

\newcommand\totalpost{5.3K}
\newcommand\totalcomment{18.5K}
\newcommand\totallike{329K}
\newcommand\totalpublishers{2.5K}
\newcommand\githuburladdress{\url{https://github.com/kooshazarei/COVID-19-InstaPostIDs}}

\newcounter{RZNumberOfComments}
\stepcounter{RZNumberOfComments}

\begin{document}

\title{A First Instagram Dataset on COVID-19\\}

\vspace{-0.4cm}
\author{\IEEEauthorblockN{Koosha Zarei\IEEEauthorrefmark{1}, Reza Farahbakhsh\IEEEauthorrefmark{1}, No\"{e}l Crespi\IEEEauthorrefmark{1}, Gareth Tyson\IEEEauthorrefmark{2}}
\IEEEauthorblockA{\IEEEauthorrefmark{1}\textit{CNRS Lab UMR5157,  T\'el\'ecom SudParis, Institut Polytechnique de Paris,} \textit{Evry, France}.\\
\{koosha.zarei, reza.farahbakhsh, noel.crespi\}@telecom-sudparis.eu}
\IEEEauthorrefmark{2}\textit{Queen Mary University of London, United Kingdom}. gareth.tyson@qmul.ac.uk
}
\vspace{-0.4cm}


\maketitle

\begin{abstract}

The novel coronavirus (COVID-19) pandemic outbreak is drastically shaping and reshaping many aspects of our life, with a huge impact on our social life. In this era of lockdown policies in most of the major cities around the world, we see a huge increase in people and professionals’ engagement in social media. Social media is playing an important role in news propagation as well as keeping people in contact.
At the same time, this source is both a blessing and a curse as the coronavirus infodemic has become a major concern, and is already a topic that needs special attention and further research. 
In this paper, we provide a multilingual coronavirus (COVID-19) Instagram dataset that we have been continuously collected since March 30, 2020. We are making our dataset available to the research community at \githuburladdress{}. We believe that this contribution will help the community to better understand the dynamics behind this phenomenon in Instagram, as one of the major social media. 
This dataset could also help study the propagation of misinformation related to this outbreak. 
\end{abstract}


\section{Introduction}
The novel coronavirus (COVID-19) was declared a pandemic by the World Health Organisation (WHO) on 11 March 2020.\footnote{\url{https://tinyurl.com/WHOPandemicAnnouncement}} 
Since then the world has experienced almost 3 million cases. To mitigate its spread, many government have therefore imposed unprecedented social distancing measures that have led to millions become housebound. 
This has resulted in a flurry of research activity surrounding both understanding and countering the outbreak~\cite{Gareth}.

As part of this, social media has become a vital tool in disseminating public health information and maintaining connectivity amongst people. Several recent studies have relied on Twitter data to better understand this~\cite{chen2020covid19,alqurashi2020large,lopez2020understanding,sharma2020coronavirus}.
These have primarily focused on health related (mis)information, but there have also been studies into online hate~\cite{schild2020go}. Despite this, there has been only limited exploration of other social modalities, such as image content.

We argue this represents a limitation, particularly considering the importance of image-based content in the dissemination of news (and misinformation)~\cite{zannettou2018origins,zannettou2017web}.
This paper introduces a COVID-19 Instagram dataset, which we make available for the research community. We have gathered data between January 5 and March 30 2020 (\S\ref{data_collection}). The dataset covers \textbf{\totalcomment{}} comments and \textbf{\totallike{}} likes from \textbf{\totalpost{}} posts. These posts have been distributed by \textbf{\totalpublishers{}} publishers. The data predominantly covers English language posts, and we provide a number of important features covering both the content and the publisher (\S\ref{sec:data_description}).
We hope that this dataset can help support a number of use cases. Hence, we conclude the paper by highlighting a number of potential uses related to COVID-19 social media analysis (\S\ref{research_topic}). Details of how to access the data is presented in \S\ref{sec:data_access}.


\section{Related Work}
\label{sec:related}
Most related to this work is the set of COVID-19 social media datasets recently released. To date, this predominantly covered textual data (\eg Twitter). To assist in this, Kazemi et al. \cite{kazemi2020english} provide a toolbox for processing textual data related to COVID-19. In terms of data, the first efforts in this direction was from authors in \cite{chen2020covid19} which provide a large Twitter dataset related to Coronavirus (by crawling major hashtags and trusted accounts). 
Another similar study \cite{alqurashi2020large}, provides an arabic Twitter dataset with a similar data collection methodology. Lopez et al.  \cite{lopez2020understanding} provide another Twitter dataset including the geolocated tweets. There are some further efforts on providing similar datasets from twitter \cite{BayesForDays, smith_kaggle,Jacobs20}.
Sharma et al. \cite{sharma2020coronavirus} also made a public dashboard\footnote{\url{https://usc-melady.github.io/COVID-19-Tweet-Analysis/}} available summarising data across more than 5 million real-time tweets. 

These Twitter datasets are being used for various use cases. For example, Saire and Navarro \cite{saire2020people} use the data to show the epidemiological impact of COVID-19 on press publications.
Singh et al. \cite{singh2020first} are also monitoring the flow of (mis)information flow across 2.7M tweets, and correlating it with infection rates to find that misinformation and myths are discussed, but at lower volume than other conversations. 
To the best of our knowledge, the only paper that has covered Instagram is by Cinelli et al. \cite{cinelli2020COVID}, who analyse Twitter, Instagram, YouTube, Reddit and Gab data about COVID-19. We complement this by making a public Instagram dataset available to the community. We redirect readers to \cite{Gareth} for a comprehensive survey of ongoing data science research related to COVID-19.

\section{Data Collection}
\label{data_collection}

We have collected public posts from Instagram by crawling all posts associated with a set of COVID-19 hashtags presented in Table \ref{tbl_dataset_tracked_hashtags}. 

\begin{table}[htbp]
\centering
\caption{Tracked Hashtags on Instagram - Release v1.0}
\vspace{-0.1cm}
\label{tbl_dataset_tracked_hashtags}
\scalebox{1.0}{
\begin{tabular}{l|rr}
\textit{\textbf{Hashtag}} & \textit{\textbf{Number of Posts}} & \textit{\textbf{Crawled Since}} \\ \hline
\textit{\#coronavirus} & 4.4K & January 5, 2020 \\
\textit{\#covid19/covid\_19} & 1.5K & January 15, 2020 \\
\textit{\#corona} & 1.0K & January 19, 2020 \\
\textit{\#stayhome} & 537 & January 30, 2020
\end{tabular}
}
\vspace{-0.3cm}
\end{table}

\pb{Methodology.} To be able to collect Instagram public content (in the shape of post), we use the official Instagram APIs \cite{InstagramAPI_online}. In particular, to get posts that are tagged with specific hashtags, the Instagram Hashtag Engine is used \cite{InstagramAPIHashtag:online}. This API returns public photos and videos that have been tagged with particular hashtags. 
MongoDB is used as the core database and data is stored as JSON records. The crawler is responsible for gathering both posts and reactions. A reaction can be active (comment) or passive (like). As it is infeasible to collect all reactions, in this dataset, we define a limitat of 500 comments and 500 likes per post. Our crawler is running on several virtual machines in parallel 24/7.
Note that we do not manually filter any posts and therefore we gather all posts containing the hashtags, regardless of the specific topics discussed within. 
The complete architecture of our crawler is described in this paper \cite{kooshaDeep}.

\pb{Release v1.0 (April 20, 2020).} 
The first version of this data collection process started on January 5, 2020 and continued until March 30, 2020. The data gathering is still running as the lockdown has not been finished in many countries around the world (at the time of writing this paper). 
During this time \textbf{\totalcomment{}} comments and \textbf{\totallike{}} likes from \textbf{\totalpost{}} public posts have been collected. These posts are distributed by \textbf{\totalpublishers{}} publishers.

\pb{Ethics.} In line with Instagram policies as well as user privacy, we only gather publicly available data that is obtainable from Instagram. 

\section{Dataset Description}
\label{sec:data_description}

To provide context for potential users of our dataset, we next brifely summarise the dataset and describe the characteristics of the content.

\begin{figure}[htbp]
\begin{center}
\vspace{-0.25cm}
  \subfloat{\includegraphics[width=1\linewidth]{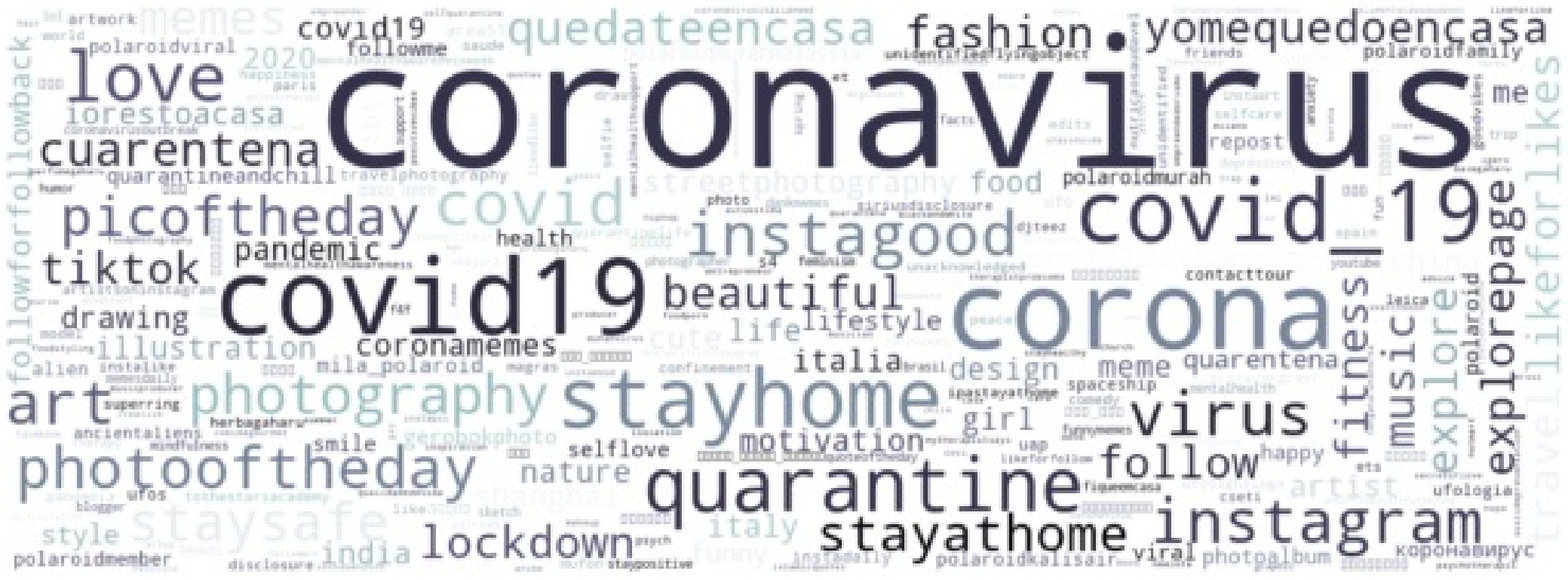}}
 \vspace{-0.1cm}
 \end{center}
\caption{Wordcloud of the most related hashtags in our dataset.}
\label{fig_hashtag_wc}
\vspace{-0.1cm}
\end{figure}

\begin{figure}[t]
\begin{center}
\vspace{-0.25cm}
  \subfloat{\includegraphics[width=1\linewidth]{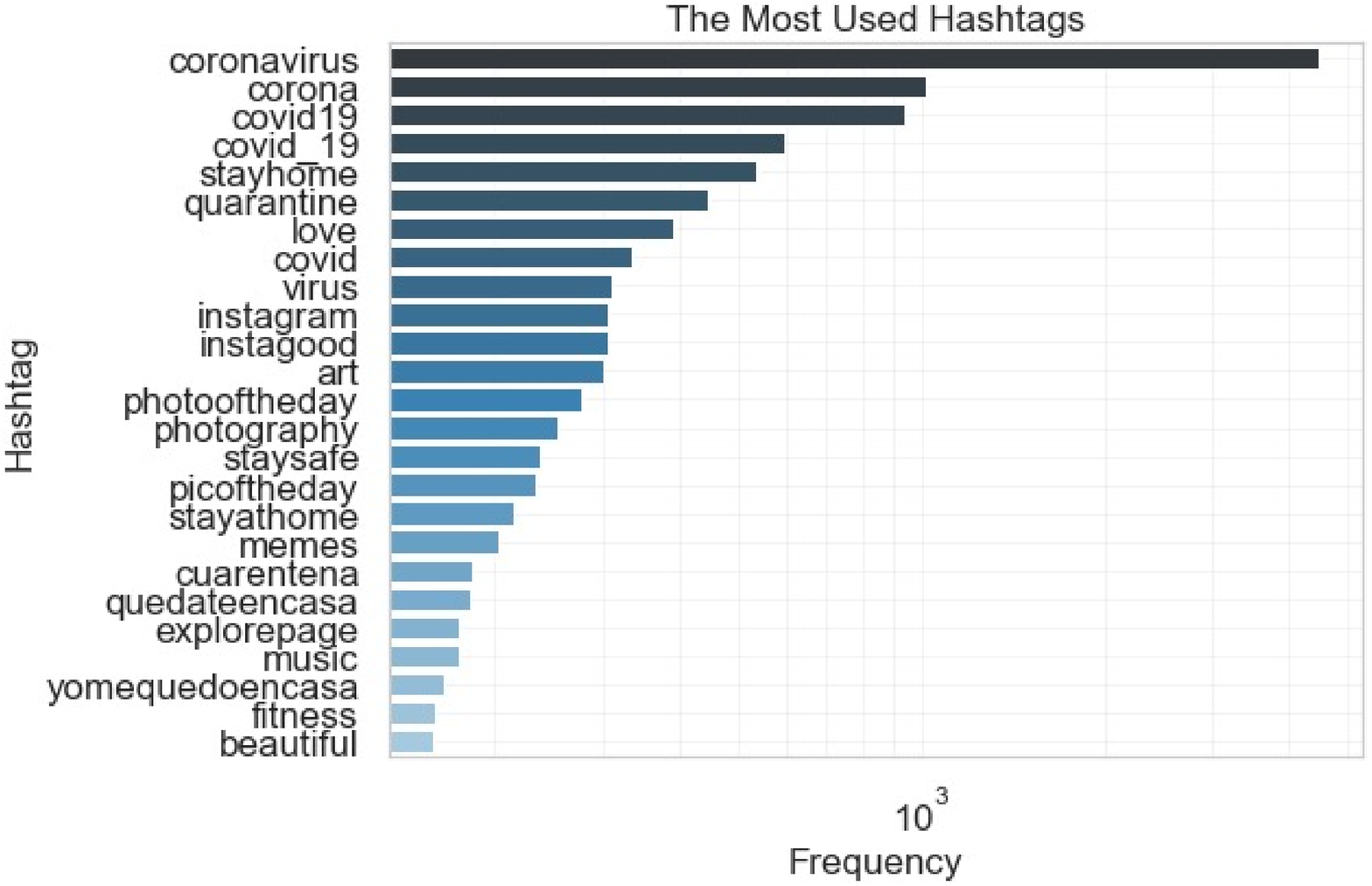}}
 \vspace{-0.1cm}
 \end{center}
\caption{Bar plot of the 25 most used hashtags.}
\label{fig_hashtag_barh}
\vspace{-0.1cm}
\end{figure}

\pb{Hashtags.} 
Recall that we gather the data by querying certain hashtags. Figure \ref{fig_hashtag_barh} presents the top hashtags tagged within the posts. Figure~\ref{fig_hashtag_wc} also presents a wordcloud of the hashtags in our dataset. This naturally includes hashtags outside of our seed set used for crawling.
There are intuitive examples, such as 
\textit{corona}, \textit{covid19}, \textit{covid\_19}, \textit{stayathome}, \textit{quarantine}, \textit{love}, \textit{covid}, \textit{virus}, and \textit{instagram}.
The are therefore the most repeated hashtags that appear with \#coronavirus. Note that this means will might miss posts that mention these concepts in other languages.

\pb{Post Language.} In order to identify the language of the post, we use spaCy library \cite{spacy2} and we apply it on the text of the caption. The language distribution is displayed in Table \ref{tbl_dataset_lang}. 
The dataset is dominated by English language content, making up almost 60\% of posts. This is driven by our choice of an English-language hashtag seed set used for data collection. That said, we have broad coverage of other widely spoken languages too, \eg Spanish (9.9\%). 
Notice that there is no official metric to determine the post language. Therefore, we highlight that this analysis could mis-classify certain posts, such as those solely hashtags or emojis.

\begin{table}[htbp]
\vspace{-0.1cm}
\centering
\caption{The Most Popular Languages}
\vspace{-0.1cm}
\label{tbl_dataset_lang}
\scalebox{1.1}{
\begin{tabular}{l|lrr}
\textit{\textbf{language}} & \textit{\textbf{code}} & \multicolumn{1}{l}{\textit{\textbf{of. \#post}}} & \multicolumn{1}{l}{\textit{\textbf{total \%}}} \\ \hline
\textit{English} & en & 3.1K & 58.3\% \\
\textit{Spanish} & es & 530 & 9.9\% \\
\textit{Portuguese} & pt & 378 & 7.1\% \\
\textit{Italian} & it & 199 & 3.7\% \\
\textit{French} & fr & 120 & 2.2\% \\
\textit{Russian} & ru & 98 & 1.8\% \\
\textit{Farsi} & fa & 96 & 1.8\% \\
\textit{Arabic} & ar & 79 & 1.4\% \\
\textit{Turkish} & tr & 68 & 1.2\% \\
\textit{Other \& non-detected} & - & 643 & 12.1\% 
\vspace{-0.1cm}
\end{tabular}
}

\end{table}

\pb{Features.} To keep data organized, the dataset is divided into four parts: \one~post content, \two~publisher information, \three~comment metrics, and \four~like features (Table \ref{table_feature_list}).
Posts contain key attributes such as a caption, list of hashtags, image/video, number of likes, number of comments, location, date, tagged list, \etc
A post is published by a public account (or public Instagram page) and in our dataset, it can be \textit{individual}, \textit{fan page}, \textit{news agency}, \textit{influencer}, \textit{blogger},  \etc
Each post receives reactions in the form of comment and like that are issued by the audience/followers.
The full feature list is presented in Table \ref{table_feature_list}. Furthermore, Table \ref{table_feature_post} presents Post and Profile characteristics in detail.

\begin{table}[htbp]
\centering
\caption{Summary of the Feature Set - Release v1.0}

\label{table_feature_list}
\scalebox{1.0}{
\begin{tabular}{l|l}
\textit{\textbf{Category}} & \multicolumn{1}{c}{\textit{\textbf{Features}}} \\ \hline
\textit{\textbf{Post}} & \textit{\begin{tabular}[c]{@{}l@{}}caption (text), caption\_language (text), \\ shortcode (number), thumbnail (binary image), \\ is\_video (bool), video\_url (text), \\ viwer\_has\_liked (bool), location (tuple), \\ hashtag (list), tagged (list), \\ mentioned (list), date\end{tabular}} \\ \hline
\textit{\textbf{\begin{tabular}[c]{@{}l@{}}Publisher \\ Profile\end{tabular}}} & \textit{\begin{tabular}[c]{@{}l@{}}username (text), id (number), \\ follower (number), followee (number), \\ media\_count (number), biography (text), \\ full name (text), is\_verified (bool), \\ is\_private (bool), profile\_picture\_url (text)\end{tabular}} \\ \hline
\textit{\textbf{Like}} & \textit{user\_id (number), username (text)} \\ \hline
\textit{\textbf{Comment}} & \textit{text, date, username (text), user\_id (number)}
\end{tabular}
}
\vspace{-0.2cm}
\end{table}

\begin{table}[h]
\centering
\caption{Post \& Profile Characteristics}
\label{table_feature_post}
\scalebox{0.94}{
\begin{tabular}{lr|lr}
\multicolumn{2}{c}{\textit{\textbf{Post}}} & \multicolumn{2}{c}{\textit{\textbf{Profile}}} \\ \hline
\textit{\textbf{name}} & \multicolumn{1}{l|}{\textit{\textbf{value}}} & \textit{\textbf{name}} & \multicolumn{1}{l}{\textit{\textbf{value}}} \\ \hline
\textit{avg. caption len} & 388 & \textit{avg. follower} & 2.6K \\
\textit{avg. received like} & 106 & \textit{avg. followee} & 925 \\
\textit{avg. received comment} & 7 & \textit{avg. mediacount} & 385 \\
\textit{is video (\%)} & 0.2\% & \textit{\begin{tabular}[c]{@{}l@{}}avg. biography len \\ (char)\end{tabular}} & 94 \\
\textit{avg. hashtag} & 16 & \textit{unverified (\%)} & 99\% \\
\textit{avg. mention account} & 0.6 & \textit{unique profiles} & 2.5K \\
\textit{avg. tagged account} & 1 & \textit{} &  \\
\textit{has location (\%)} & 1\% &  & 
\end{tabular}
}
\vspace{-0.3cm}
\end{table}


\section{Potential Research Topics}
\label{research_topic}

We hope that the dataset can support diverse research activities. Below we list a subset of potential topics, we believe the dataset could support: 

\begin{enumerate}
\item \emph{Fake news, misinformation and rumors spreading:}. Several researcher have started to inspect COVID-19 misinformation. As an example, an infodemic observatory have analyzed more than 100M public messages to understand the digital response in online social media to COVID-19 outbreak.\url{https://covid19obs.fbk.eu} In another study, Sharma et al. \cite{sharma2020coronavirus} made a public dashboard available summarising data from real-time tweets in
in \url{https://usc-melady.github.io/COVID-19-Tweet-Analysis/} with a focus to misinformation spread analysis. We believe that our Instagram data could be used to evaluate the flow of misinformation (\eg memes) on Instagram.

\item \emph{Bot Population and bot generated content:} It is well known that bot content plays a prominent role in social media data~\cite{gilani2019large}. These have the capacity of amplify misinformation or even act against public health policies (\eg encourgaging a breakdown in social distancing). We posit that the data could be used to explore the role of bots in this dissemination.

\item \emph{Behavioral change analysis during the pandemic:} The social distancing measures are created an unprecedented change to millions of people's lives. Understanding the behavioral consequences of this is vital for understanding things like adherence to social distancing policies and mental health consequences.

\item \emph{Information sharing related Covid-19:} Information flow is vital during periods of emergency. We posit that the dataset can be used to understand the flow of information, as well as people's reactions to such information. 

\end{enumerate}

\section{Dataset Access}
\label{sec:data_access}

The presented dataset is accessible in this address on Github platform: \githuburladdress{}. This is the first version of the dataset and we are still collecting data. Hence, we hope to make further versions available in the coming weeks and months.
We publish this dataset in agreement with Instagram's Terms \& Conditions \cite{InstagramDataPolicy:online}, and as it is not possible to release the post content and reactions, we just distribute the post ID's. These are known as the \textit{shortcodes}. 
Researchers can then simply retrieve post content through IDs by the help of some open-source projects such as Instaloader \cite{instaloaderGithub} that have been developed for such purposes.
For any further question, please contact Koosha Zarei at \textit{koosha.zarei@telecom-sudparis.eu}.


\bibliographystyle{unsrt}
\bibliography{paper}

\begin{thebibliography}{10}

\bibitem{Gareth}
Siddique Latif, Muhammad Usman, Sanaullah Manzoor, Waleed Iqbal, Junaid Qadir,
  Gareth Tyson, Ignacio Castro, Adeel Razi, Maged N~Kamel Boulos, and Jon
  Crowcrroft.
\newblock Leveraging data science to combat covid-19: {A} comprehensive review.
\newblock
  \url{http://www.eecs.qmul.ac.uk/~tysong/files/COVID_19_Review_v1.pdf}, 2020.

\bibitem{chen2020covid19}
Emily Chen, Kristina Lerman, and Emilio Ferrara.
\newblock Covid-19: The first public coronavirus twitter dataset, 2020.

\bibitem{alqurashi2020large}
Sarah Alqurashi, Ahmad Alhindi, and Eisa Alanazi.
\newblock Large arabic twitter dataset on covid-19, 2020.

\bibitem{lopez2020understanding}
Christian~E. Lopez, Malolan Vasu, and Caleb Gallemore.
\newblock Understanding the perception of covid-19 policies by mining a
  multilanguage twitter dataset, 2020.

\bibitem{sharma2020coronavirus}
Karishma Sharma, Sungyong Seo, Chuizheng Meng, Sirisha Rambhatla, Aastha Dua,
  and Yan Liu.
\newblock Coronavirus on social media: Analyzing misinformation in {Twitter}
  conversations.
\newblock {\em arXiv preprint arXiv:2003.12309}, 2020.

\bibitem{schild2020go}
Leonard Schild, Chen Ling, Jeremy Blackburn, Gianluca Stringhini, Yang Zhang,
  and Savvas Zannettou.
\newblock " go eat a bat, chang!": An early look on the emergence of sinophobic
  behavior on web communities in the face of covid-19.
\newblock {\em arXiv preprint arXiv:2004.04046}, 2020.

\bibitem{zannettou2018origins}
Savvas Zannettou, Tristan Caulfield, Jeremy Blackburn, Emiliano De~Cristofaro,
  Michael Sirivianos, Gianluca Stringhini, and Guillermo Suarez-Tangil.
\newblock On the origins of memes by means of fringe web communities.
\newblock In {\em Proceedings of the Internet Measurement Conference 2018},
  pages 188--202, 2018.

\bibitem{zannettou2017web}
Savvas Zannettou, Tristan Caulfield, Emiliano De~Cristofaro, Nicolas
  Kourtelris, Ilias Leontiadis, Michael Sirivianos, Gianluca Stringhini, and
  Jeremy Blackburn.
\newblock The web centipede: understanding how web communities influence each
  other through the lens of mainstream and alternative news sources.
\newblock In {\em Proceedings of IMC}, 2017.

\bibitem{kazemi2020english}
Salma Kazemi~Rashed, Johan Frid, and Sonja Aits.
\newblock English dictionaries, gold and silver standard corpora for biomedical
  natural language processing related to {SARS-CoV-2} and {COVID-19}.
\newblock {\em arXiv}, pages arXiv--2003, 2020.

\bibitem{BayesForDays}
C.~Jacobs.
\newblock Coronada: Tweets about covid-19.
\newblock https://github.com/BayesForDays/coronada, 2020.

\bibitem{smith_kaggle}
Smith, “coronavirus (covid19) tweets”, mar 2020. [online]. available:
  \url{www.kaggle.com/smid80/coronavirus-covid19-tweets}.

\bibitem{Jacobs20}
Cassandra Jacobs.
\newblock Coronada, 2020.

\bibitem{saire2020people}
Josimar E~Chire Saire and Roberto~C Navarro.
\newblock What is the people posting about symptoms related to {Coronavirus} in
  {Bogota, Colombia?}
\newblock {\em arXiv preprint arXiv:2003.11159}, 2020.

\bibitem{singh2020first}
Lisa Singh, Shweta Bansal, Leticia Bode, Ceren Budak, Guangqing Chi, Kornraphop
  Kawintiranon, Colton Padden, Rebecca Vanarsdall, Emily Vraga, and Yanchen
  Wang.
\newblock A first look at {COVID-19 information and misinformation sharing on
  Twitter}.
\newblock {\em arXiv preprint arXiv:2003.13907}, 2020.

\bibitem{cinelli2020COVID}
Matteo Cinelli, Walter Quattrociocchi, Alessandro Galeazzi, Carlo~Michele
  Valensise, Emanuele Brugnoli, Ana~Lucia Schmidt, Paola Zola, Fabiana Zollo,
  and Antonio Scala.
\newblock The {COVID-19} social media infodemic.
\newblock {\em arXiv preprint arXiv:2003.05004}, 2020.

\bibitem{InstagramAPI_online}
Instagram.
\newblock Official api graph instagram.
\newblock https://developers.facebook.com/docs/instagram-api, January 2020.

\bibitem{InstagramAPIHashtag:online}
Instagram.
\newblock Instagram hashtag search.
\newblock
  https://developers.facebook.com/docs/instagram-api/guides/hashtag-search,
  February 2020.

\bibitem{kooshaDeep}
Koosha Zarei, Reza Farahbakhsh, and Noel Crespi.
\newblock Deep dive on politician impersonating accounts in social media.
\newblock In {\em 2019 IEEE Symposium on Computers and Communications (ISCC)
  (IEEE ISCC 2019)}, Barcelona, Spain, June 2019.

\bibitem{spacy2}
Matthew Honnibal and Ines Montani.
\newblock {spaCy 2}: Natural language understanding with {B}loom embeddings,
  convolutional neural networks and incremental parsing.
\newblock To appear, 2017.

\bibitem{gilani2019large}
Zafar Gilani, Reza Farahbakhsh, Gareth Tyson, and Jon Crowcroft.
\newblock A large-scale behavioural analysis of bots and humans on twitter.
\newblock {\em ACM Transactions on the Web (TWEB)}, 13(1):1--23, 2019.

\bibitem{InstagramDataPolicy:online}
Data Policy.
\newblock Instagram data policy.
\newblock https://help.instagram.com/519522125107875, March 2020.

\bibitem{instaloaderGithub}
Instaloader.
\newblock Instaloader.
\newblock https://github.com/instaloader/instaloader, January 2020.

\end{thebibliography}

\end{document}